\documentclass[11pt]{article}

\usepackage{amsfonts}
\usepackage{amsmath,amssymb,epsf,indentfirst}
\usepackage{graphicx}
\usepackage{verbatim}
\usepackage{cite}
\usepackage[usenames,dvipsnames]{xcolor}

\numberwithin{equation}{section}

\makeatletter
\@addtoreset{equation}{section}
\makeatother

\def\arXivid{arXiv:}
\def\hepth{hep-th/}
\def\None{\mathcal{N}\,{=}\,1}
\def\Ntwo{\mathcal{N}\,{=}\,2}
\def\Nthree{\mathcal{N}\,{=}\,3}
\def\Nfour{\mathcal{N}\,{=}\,4}
\def\Nsix{\mathcal{N}\,{=}\,6}
\def\Neight{\mathcal{N}\,{=}\,8}

\def\Nf{N_\text{f}}
\def\nf{n_\text{f}}

\newcommand\plb[3]   {
		{{\it Phys.\ Lett.\ }{\bf B #1} (#2) #3}}
\newcommand\ijmpa[3] {
		{{\it Int.\ J.\ Mod.\ Phys.\ }{\bf A #1} (#2) #3}}

\newcommand\cqg[3]   {
		{{\it Class.\ and Quant.\ Grav.\ }{\bf #1} (#2) #3}}
\newcommand\cmp[3]   {
		{{\it Commun.\ Math.\ Phys.\ }{\bf #1} (#2) #3}}

\newcommand\atmp[3] {
		{{\it Adv.\ Theor.\ Math.\ Phys.\ }{\bf #1} (#2) #3}}
\newcommand\npb[3]    {
		{{\it Nucl.\ Phys.\ }{\bf B #1} (#2) #3}}
\newcommand\prl[3]   {
		{{\it Phys.\ Rev.\ Lett.\ }{\bf #1} (#2) #3}}
\newcommand\prd[3]   {
		{{\it Phys.\ Rev.\ }{\bf D #1} (#2) #3}}
\newcommand\jhep[3]  {
		{{\it J. High Energy Phys.\ }{\bf #1} (#2) #3}}
\newcommand\ptp[3]   {
		{{\it Prog.\ Theor.\ Phys.\ }{\bf #1} (#2) #3}}
\newcommand{\Z}{\mathbb{Z}}
\newcommand{\Slash}[1]{{\ooalign{\hfil$#1$\hfil\crcr\raise.167ex\hbox{/}}}}
\newcommand{\SL}{\Slash{L}}

\newcommand{\x}{{\times }}

\begin{document}
\begin{titlepage}

\begin{flushright}
IPMU13-0090
\end{flushright}

\begin{center}
\noindent{\Large \textbf{Dualities from large \boldmath $N$ orbifold equivalence in
		Chern-Simons-matter theories with flavor}}\\
\vspace{15mm} Mitsutoshi Fujita\footnote{mf29@uw.edu}
\vspace{1cm}

{\it Department of Physics, University of Washington, Seattle, Washington 98195-1560, USA} \\ 
{\it and} \\ 
{\it Kavli Institute for the Physics and Mathematics of the Universe (WPI), University of Tokyo, Kashiwa, Chiba 277-8583, Japan} \\

\vskip 3em

\begin{abstract}
{ We study large-$N$ orbifold equivalences involving
    three-dimensional $\Nthree$ and $\Nfour$
    supersymmetric quiver Chern-Simons matter theories.
    The gravity dual of
    the $\Nthree$ Chern-Simons matter theory is described
    by $AdS_4\times \mathcal{M}_7$, where the tri-Sasaki manifold
    $\mathcal{M}_7$ is known as the Eschenburg space.  We find evidence that
    a large-$N$ orbifold equivalence for the $\Nfour$ case continues from
    the M-theory limit to the weak coupling limit. For the $\Nthree$ case,
    we find consistent large-$N$ equivalences involving
    a projection changing the
    nodes of the gauge groups, and also for a projection changing
    Chern-Simons levels where {for the latter projection, the BPS monopole operators behave as expected in large-$N$ 
    equivalence.}    
    For both cases we show, using the
    gravity dual, that the critical temperature of the confinement/deconfinement
    transition does not change and the entropy behaves as expected under
    the orbifold equivalence. We show that large-$N$ orbifold equivalence changing Chern-Simons levels can be explained using the planar equivalence in the mirror dual. }

\end{abstract}
\end{center}
\end{titlepage}

\section{INTRODUCTION}

A variety of nonperturbative equivalences are known to relate the
large-$N$ limits of many non-Abelian gauge theories (where $N$ is the
rank of the gauge group).
These include, for example,
equivalences relating $SU(N)$ and $SO(N)$ or $Sp(N)$ theories~\cite{Lovelace:1982hz,Cherman:2010jj,Hanada:2011ju},
equivalences relating theories with adjoint representation matter fields
to theories with bifundamental or other tensor representation matter fields~\cite{Kachru:1998ys,Lawrence:1998ja,Bershadsky:1998mb,Armoni:2003gp},
and equivalences relating toroidally compactified theories with differing
compactification radii~\cite{Eguchi:1982nm}. 
Underlying these, and other, examples of large-$N$ equivalences is a network
of orbifold projections.
In field theories, an orbifold projection is a mapping which, given
an initial (``parent'') theory, plus some chosen discrete symmetry
of this theory, constructs a new (``daughter'') theory by removing
from the parent theory all degrees of freedom which are not invariant
under the chosen discrete symmetry.
In suitable cases, observables in the parent and daughter theories which are
invariant under appropriate symmetries coincide in the large-$N$ limit~\cite{Kovtun:2004bz,Kovtun:2005kh}.

However, not all orbifold projections lead to large-$N$ equivalences.%
\footnote
    {%
    For example, a projection defined by a
    $\Z_2$ subgroup of the gauge symmetry, combined with $(-1)^F$,
    maps $U((p{+}q)N)$ $\None$ supersymmetric Yang-Mills theory to a
    $U(p N) \times U(q N)$ quiver theory with bifundamental fermions.
    Only in the special case of $p = q$ are the parent and daughter
    theories related by a large-$N$ equivalence.
    Nonequivalence for $p \ne q$ may be easily confirmed by
    considering the high-temperature thermodynamics of these theories.
    }
Valid large-$N$ equivalences appear to be associated with ``invertible''
orbifold projections~\cite{Kovtun:2007py}, that is, cases where some orbifold projection
maps theory $A$ to theory $B$ while a different projection maps theory
$B$ back to theory $A$ (with a smaller gauge group rank).
This is an empirical observation based on examining numerous cases,
but we are unaware of any general proof.
In light of this, it is interesting to consider more diverse examples
of orbifold projections and investigate when and if they lead to valid
large-$N$ equivalences.

In this paper, we discuss several orbifold projections of
$d = 3$, $\mathcal N = 6$ ABJM theory, as well as
generalizations of ABJM theory which include fundamental
representation flavors. The first case we consider is a $\Z_p$ projection which
relates the $\Nsix$ $U(pN)_{kp} \times U(pN)_{-kp}$
ABJM theory to an $\mathcal N=4$ $[U(N)_k \times U(N)_{-k}]^p$
 quiver Chern-Simons matter theory with a $\Z_p$ global symmetry. To see unbroken $\Z_p$ global symmetry in the weak coupling limit, we confirm large-$N$ orbifold equivalence using the free theory on $S^1\times S^2$ and computing the free energy, the Polyakov loop vacuum expectation value (VEV) along the time direction, and the critical temperature of the confinement/deconfinement transition as analyzed similarly to show large-$N$ equivalences of QFTs on $S^1\times S^3$~\cite{Larsen:2007bm,Unsal:2007fb}. We also show large-$N$ orbifold equivalence using the AdS/CFT correspondence~\cite{Maldacena:1997re,Gubser:1998bc,Witten:1998qj} in both the type-IIA string theory and the outside of the planar limit of the M theory where $k$ is fixed and $N$ is taken to be large~\cite{Kiritsis:2010xc} (see also Ref.~\cite{Armoni:2008kr}). 

We then add $\Nf$ fundamental representation hypermultiplets to 
these theories. This turns the parent ABJM theory into an
$\Nthree$ $U(pN)_{kp} \times U(pN)_{-kp}$ quiver Chern-Simons matter theory.
Provided $\Nf$ is divisible by $p$, $\Nf \equiv p \nf$,
one may define an orbifold projection mapping theory to an
$\Nthree$ $[U(N)_k \times U(N)_{-k}]^p$ quiver Chern-Simons matter theory
with $\nf$ fundamentals at each node and a $\Z_p$ global symmetry.
To obtain observables in the neutral sector of $\mathbb{Z}_p$ symmetry, we perform the projection of the mesonic operators. We also show large-$N$ orbifold equivalence using the gravity dual in both the type-IIA string theory and the outside of the planar limit in the M theory. The gravity dual of an
$\Nthree$ $[U(N)_k \times U(N)_{-k}]^p$ quiver Chern-Simons matter theory
with fundamentals has been studied in Refs.~\cite{Hohenegger:2009as,Gaiotto:2009tk,Hikida:2009tp,Fujita:2009xz} and
for massive flavors in Refs.~\cite{Jensen:2010vx,Zafrir:2012yg}.\footnote{There
are other $\mathcal{N}\le 2$ Chern-Simons matter theories including flavor, in which
flavor are D6 branes in the type-IIA string theory~\cite{Jafferis:2009th,Benini:2009qs,Conde:2011sw}.} The gravity dual of Chern-Simons matter theory with backreacted flavor has been proposed as the $d=11$ supergravity on $AdS_4\times
\mathcal{M}_7$  where flavor becomes the Kaluza-Klein magnetic monopole as a soliton.\footnote{The cone over tri-Sasaki $\mathcal{M}_7$ is
given by a $d=8$ toric hyper-K{\"a}hler manifold with $Sp(2)$ holonomy
and with 3/16 supersymmetry~\cite{GGPT}.}
Here, tri-Sasaki $\mathcal{M}_7$
is given by the Eschenburg space~\cite{E,1,2,3} parametrized by
three relatively prime numbers $(t_1,t_2,t_3)$, which are read off according to the number and charge of 5-branes in the dual type-IIB elliptic brane configuration.

Finally, we consider some level-changing projection which is not understood well. 
 Large-$N$ equivalence between ABJM theories in terms of the level-changing projection was analyzed in Refs.~\cite{Hanada:2011yz,Hanada:2011zx} based on the proof using the AdS/CFT correspondence in M theory. Since we need proof in terms of the string theory and mirror symmetry to understand the level-changing projection, it is interesting to study the level-changing projection further. As a generalization, we apply the level-changing projection for $d=3$, $\mathcal{N}=3$ Chern-Simons matter theories with flavor. We analyze the $AdS_4\times
\mathcal{M}_7$ gravity dual in large-$N$ orbifold equivalence and show that the curvature radius of the two equivalent theories becomes the same. Since $(t_1,t_2,t_3)$ depend on the discrete number of the orbifold, the equivalence in terms of the level-changing projection is nontrivial. We analyze the BPS monopole operators and Bekenstein-Hawking entropy in large-$N$ equivalence. Moreover, we analyze the critical temperature of the confinement/deconfinement transition using the entropy. When the discrete symmetry of the level-changing projection is unbroken, we can observe large-$N$ orbifold equivalence in the phase transition. We also show that  the planar dominance holds in the very strong coupling limit of the M theory since the equivalence holds even outside the 't Hooft limit, as long as there exists a classical gravity dual~\cite{Fujita:2012cf,Azeyanagi:2012xj}.

The content of this paper is as follows. In Sec. II, we review
the $\Nsix$ ABJM theory and $\Nfour$ Chern-Simons matter
theories and show that the orbifold projection of the $\Nsix$
ABJM theory gives a $\Nfour$ Chern-Simons matter theory.
After that, we introduce flavor in the Chern-Simons matter
theory and discuss the orbifold equivalence in the presence of flavor. 
In Sec. III, we consider the free Chern-Simons matter 
theory on $S^1\times S^2$ at finite temperature and analyze large-$N$
orbifold equivalence. In Sec. IV, we analyze large-$N$ equivalence in the type-IIA gravity dual to the Chern-Simons matter theory with small amounts of flavor by introducing probe D6 branes. In Sec. V, we analyze the orbifold equivalence between the $\Nthree$ Chern-Simons matter theories using the gravity 
dual with backreacted flavor. As observables invariant under chosen discrete symmetry, we analyze BPS monopole operators in large-$N$ orbifold equivalence. We also compare the Bekenstein-Hawking entropy and the 
Hawking-Page transition between the parent theory and the daughter
theory under the orbifold. In Sec. VI, we explain the orbifold
changing Chern-Simons levels by using the mirror symmetry of the
type-IIB elliptic D3-brane system. It turns out that such an orbifold
changes the number of nodes and NS5 branes in the mirror theory
side.

\section{$\Nfour$ CHERN-SIMONS MATTER THEORY}

\begin{figure}[htbp]
   \begin{center}
     \includegraphics[height=4cm]{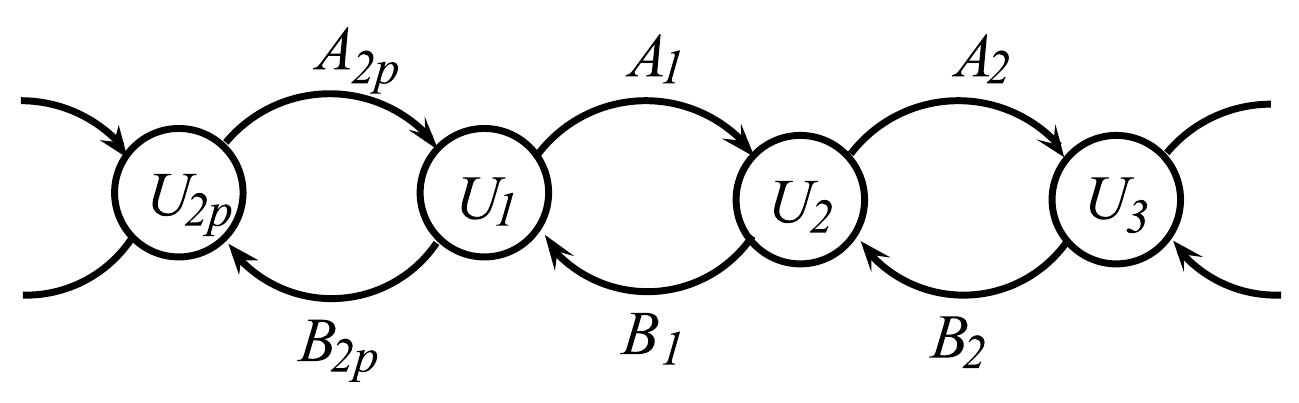}
   \end{center}
   \caption[quiverzu]{A part of the quiver diagram of $\Nfour$ Chern-Simons matter theory. $U_I$ represents the $I$ th node. $A_I,B_I$ show chiral multiplets composing the $\Nfour$ (twisted) hypermultiplet.}
\label{fig:quiverzu}
\end{figure} 

In this section, we briefly review the $\Nfour$ Chern-Simons matter theory proposed in Refs.~\cite{Imamura:2008nn,H,I}. 
The field content in the $d=3$ $\Nfour$ Chern-Simons theory becomes the $\Nfour$ vector multiplet $(V_I,\Phi_I)$, $\Nfour$ hypermultiplet $(A_I,B_I)$, and 
$\Nfour$ twisted hypermultiplet $(A_{I}',B_{I}')$. 
Here, the chiral multiplets $A_I,A_I'$ and $\bar{B}_I,\bar{B}_I'$ transform under
the gauge symmetry $\prod_{I=1}^{2p} U(N)_I$ with $U(N)_{2p+1}=U(N)_1$
as (${\mathbf N}_I$,$\overline{{\mathbf N}}_{I+1}$). In terms of
components, $A_I=(h_I,\psi_{I, \alpha})$,
$B_I=(\tilde{h}_I,\tilde{\psi}_{I, \alpha})$,
$A_{I}'=({h}_{I}',{\psi}_{I, \alpha}')$, and
$B_{I}'=(\tilde{h}_{I}',\tilde{\psi}_{I, \alpha}')$. We consider the quiver
diagram with a cyclic configuration in Fig. \ref{fig:quiverzu}.
{We label hypermultiplets with the associated numbers $n_I$. Numbers $n_I$ are $0$ for untwisted hypermultiplets and 1 for twisted hypermultiplets.}

The Lagrangian for the hypermultiplets and the Chern-Simons terms is composed from kinetic terms and 
superpotential as follows:
\begin{align}
&S_{CS}=\sum^{2p}_{I=1}\dfrac{ik_I}{2}\Big[\int d^3xd^4\theta \int ^1_0dt \mbox{tr}\Big(V_I\bar{D}(e^{-2tV_I}De^{2tV_I})\Big) \nonumber \\
&-\mbox{tr}\Big(\int d^3xd^2\theta \Phi_I^2+c.c.\Big) \Big], \\
&S_{hyper}=-\sum^{2p}_{I=1}\int d^3xd^4\theta \mbox{tr}\Big(\bar{A}_Ie^{2V_I}A_Ie^{-2V_{I+1}}+B_Ie^{-2V_I}\bar{B}_Ie^{2V_{I+1}}\Big) \nonumber\\
&+\sum^{2p}_{I=1}\Big(\int d^3xd^2\theta \sqrt{2}i\mbox{tr}(B_I\Phi_I A_I-B_IA_I\Phi_{I+1}+c.c.)\Big), \label{HYP511}
\end{align}
where the Chern-Simons level of the $U(N)_I$ gauge group is given by 
$k_I=k(n_{I+1}-n_I)$. In this paper, we consider the particular case where 
the hypermultiplets and twisted hypermultiplets are aligned mutually, namely, $k_I=\pm k$ for even(odd) $I$, and the number of the untwisted multiplet is the same as the number of the twisted multiplet~\cite{I}.
Nondynamical adjoint fields $\Phi^I$ in F terms and auxiliary fields in the vector multiplet are integrated out.
 $SO(4)_R\sim SU(2)_t\times SU(2)_{unt}$
 $R$ symmetry rotates $(h_I,\bar{\tilde{h}}_I)$ and $(h_{I}^{\prime},\bar{\tilde{h}}_{I}^{\prime})$, respectively. There are also the baryonic $U(1)_b$ and the diagonal $U(1)_d$, which act on $(h_I,h_I')$ and $(\bar{\tilde{h}}_I,\bar{\tilde{h}}_I')$ in two different ways. These symmetries agree with the isometry $(SU(2)\times U(1))^2$ of its moduli space $(\mathbb{C}^2/\mathbb{Z}_p\times \mathbb{C}^2/\mathbb{Z}_p)/\mathbb{Z}_k$~\cite{Imamura:2008nn,Jafferis:2008qz}.

Remember that for two nodes $p=1$, the $\Nfour$
Chern-Simons matter theory becomes the  ABJM theory preserving
enhanced $\Nsix$ supersymmetry and enlarged $SU(4)_R$
$R$ symmetry for $k>2$~\cite{Aharony:2008ug}. In the ABJM theory,
two nodes are joined by four links representing chiral multiplets, 
$A_1,A_2,B_1,B_2$.
Moreover, the $U(N)\times U(N)$ ABJM theory has enhanced $\Neight$
supersymmetry for $k=1,2$~\cite{Klebanov:2009sg,Samtleben:2010eu}
(see also Ref.~\cite{Gustavsson:2009pm}). The amount of  supersymmetry
of the Chern-Simons matter theory is also reviewed in the next
section in the view of the multiple M2 brane theory on an orbifold
of $\mathbb{C}^4$.
 
\subsection{Orbifolds of the $\Nsix$ ABJM theory}

In this section, we present the review of the orbifold projection of $\Nsix$ $U(nN)_{kn}\times U(nN)_{-kn}$ ABJM action as the multiple M2-brane theory which corresponds to the $p=1$ case of the $\Nfour$ Chern-Simons matter theory~\cite{Benna:2008zy}. Here, we have quantized Chern-Simons levels in terms of $n$.   
In this quantization, 't Hooft coupling $\lambda =N/k$ is independent of $n$.
It is known that the ABJM theory is the multiple M2-brane theory placed at the $\mathbb{Z}_{nk}$ orbifold of $\mathbb{C}^4$. Note that this $\mathbb{Z}_{nk}$ orbifold does not correspond to the orbifold projection of the $U(nN)_{kn}\times U(nN)_{-kn}$ ABJM theory. 

Here, we focus on a single M2 brane instead of the multiple M2 branes and introduce the four complex variables $y^A$ to specify $\mathbb{C}^4$, the isometry of which is $SO(8)$.  The orbifold action acts on $y^A$ as $y^A\to e^{{2\pi i}/(nk)}y^A$. 
 The orbifold $\mathbb{Z}_{kn}$ action also acts on the spinors of $SO(8)$ as
\begin{align}
\epsilon = e^{2\pi i(s_1+s_2+s_3+s_4)/(kn)}\epsilon, \label{ORB22}
\end{align}
where $s_i$ $(i=1,\dots,4)$ takes the values $\pm 1/2$. The chirality condition of the spinor imposes on $s_i$ the condition that $\sum_{i=1}^4s_i$ is even, giving an eight-dimensional spinor representation. The orbifolded theory then has six spinors out of eight spinors, or $d=3$ $\Nsix$ supersymmetry for $kn>2$ ($d=3$ $\Neight$ supersymmetry for $kn=1,2$).

The $\mathbb{Z}_n$ orbifold projection preserving $\Nfour$ supersymmetry is given by
\begin{align}
y^A=e^{{2\pi i}/{n_A}}y^A,\quad (n_1,n_2,n_3,n_4)=(n,\infty,n,\infty). \label{ORB23}
\end{align}
We can show the supersymmetry by considering the action on the
spinor as $\epsilon \to e^{2\pi i(s_1+s_2)/n}\epsilon$, which leaves
four spinors.\footnote{It is known that, changing the basis of the
orbifold action, the above orbifold action is equivalent to
\begin{align}
y^A=e^{{2\pi i}/{n_A}}y^A,\quad (n_1,n_2,n_3,n_4)=(n,n,-n,-n). \label{ORB24}
\end{align}
This orbifold Eq.~\eqref{ORB24} also preserves $\Nfour$ supersymmetry.}

Secondly, we consider the multiple M2-brane theory. 
In the non-Abelian gauge theory, we introduce the element of the
$\mathbb{Z}_n$ orbifold projection from the element of each gauge
group $U(nN)\times U(nN)$ as
\begin{align}\label{OME26}
\Omega = \text{diag}(1_{\mathbf{N}},\ \nu 1_{\mathbf{N}},\ \nu^2 1_{\mathbf{N}},\ \dots\ ,\nu^{n-1} 1_{\mathbf{N}}),
\end{align}
where $1_{\mathbf{N}}$ is the $N\times N$ identity matrix and we
have defined the phase $\nu =e^{2\pi i/n}$.
We combine chiral multiplets of bifundamental matters into the
following multiplet, transforming under $SU(4)$ enhanced $R$ symmetry:
\begin{align}
y^A=(A_1,B_2,\bar{B}_1,\bar{A}_2).
\end{align}
The $\mathbb{Z}_n$ orbifold action~\cite{Douglas:1996sw} acts on the bifundamental fields $y^A$, $V_I$ and $\Phi_I$ ($I=1,2$) as 
\begin{align}\label{ORB30}
&y^1 =A_1= \nu\Omega A_1\Omega^{-1} ,\quad y^2 =B_2=\Omega B_2 \Omega^{-1}, \\
& \bar{y}^3=B_1= \nu^{-1}\Omega B_1\Omega^{-1},\quad \bar{y}^4=A_2= \Omega A_2\Omega^{-1}, \\
&V_I=\Omega V_I\Omega^{-1},\quad \Phi_I= \Omega \Phi_I\Omega^{-1}.
\end{align}

We show that the orbifolded theory agrees with the $\Nfour$ $[U(N)_k\times U(N)_{-k}]^n$ Chern-Simons matter theory with $p=n$ obtained in the previous section. Remember that we quantized the Chern-Simons levels of the mother theory by $n$, and the Chern-Simons levels of the daughter theory become $\pm k$ for each node~\cite{Terashima:2008ba}. We observe that the 't Hooft coupling $\lambda =N/k$ becomes the same between the mother theory and the daughter theory, which is the condition of the orbifold equivalence. 
It is known that the orbifolded theory has the moduli space $\mathbb{C}^2/\mathbb{Z}_{kn}\times \mathbb{Z}_n$.

Note that the Douglas-Moore orbifold projection is constructed~\cite{Douglas:1996sw}   not in the M-theory limit but in the type-IIA limit. This problem is resolved in Ref.~\cite{Terashima:2008ba} by moving the M2 branes from the orbifold singularity of the moduli space and by obtaining the D2 branes in the type-IIA superstring theory.

\subsection{Adding flavor to the Chern-Simons matter theory}

In this section, we review the $\Nthree$ Chern-Simons matter theory with flavor,  which can be constructed from the $d=3$ $\Nfour$ Chern-Simons matter theory by adding flavor fields~\cite{Fujita:2009xz} (see also Refs.~\cite{Hohenegger:2009as,Gaiotto:2009tk,Hikida:2009tp}). After that, we discuss the orbifold equivalence between Chern-Simons matter theories with flavor. 
We add massless flavor to the $\Nfour$ Chern-Simons matter theory, namely, $N_F^I=N^0_F$ fundamental hypermultiplets aligned evenly among the different groups $(Q_\alpha^I, \tilde{Q}_\alpha^I)$ $(\alpha=1,\cdots,N_F^0,\ I=1,\dots ,2p)$ transforming under the $I$ th gauge group as $(\mathbf{N},\bar{\mathbf{N}})$ with $\sum _I N^I_F =N_F$. 
Then, we add the D term to the $\Nfour$ action as follows:
\begin{eqnarray*}
{\mathcal S}_{flavor1} =- \mbox{Tr} \sum_{\alpha, I} 
\int d^3 x  d^4 \theta ~ (\overline{Q}_{\alpha}^{ I }e^{2V_I} 
Q_{\alpha}^I 
+ \tilde{Q}_{\alpha}^I e^{-2 V_I} \overline{\tilde{Q}}_{\alpha}^{ I} ).
\end{eqnarray*}
The potential term is also added:
\begin{align}
S_{flavor2}=-\int d^3xd^2\theta\sum_{I,\alpha}i\sqrt{2}\tilde{Q}^I_{\alpha}\Phi_IQ^I_{\alpha}+c.c.
\end{align}
The nondynamical field $\Phi_I$ can be integrated out.\footnote{After integrating out $\Phi_I$, the superpotential is rewritten as \begin{align}
\int d^3xd^2\theta\sum_{i=1}^{2p}\dfrac{i}{k_i}\mbox{tr}(A_iB_i-B_{i-1}A_{i-1}-Q^i_{\alpha}\tilde{Q}^i_{\alpha})^2.
\end{align} The convention of the superpotential is different from that of Ref.~\cite{Fujita:2009xz} by the change $A_i\to B'_{i+1}$ and $B_i\to A'_{i+1}$ (See also Ref.~\cite{Gaiotto:2009tk}).}
$R$ symmetry is now broken to $SU(2)_R$, which is the diagonal part of $SU(2)_{unt}\times SU(2)_t$, while $U(1)_b$ symmetry is unchanged. This theory also has  $SU(2)_d$ global symmetry commuting with $SU(2)_R$~\cite{Ammon:2009wc}. $SU(2)_d$ is also a subgroup of $SU(4)_R$ in the ABJM theory.

For $p=1$, the mesonic field in this theory is constructed as $\tilde{Q}_1(A_1B_1)^lQ_1$ or $\tilde{Q}_1$ $(A_1B_1)^lA_1Q_2$. If there is one flavor, the former type of the mesonic operator exists, and if there are two types of flavors with different gauge groups, both the mesonic operators exist. 

We now consider the $\mathbb{Z}_n$ orbifold projection of the Chern-Simons matter theory with flavor for $p=1$. The orbifold action for the supersymmetry multiplets except for the fundamental hypermultiplets is given in Eq.~\eqref{ORB30}.
Using the projection matrix Eq.~\eqref{OME26}, the $\mathbb{Z}_n$ orbifold action for the fundamental hypermultiplets is given by 
\begin{align}
Q^I_{\alpha}=\nu^{m^I_{\alpha}} \Omega Q^I_{\alpha},\quad \tilde{Q}^I_{\alpha}=\nu^{-m^I_{\alpha}} \tilde{Q}^I_{\alpha}\Omega^{-1},
\end{align}
where $I=1,2$ and $m^I_{\alpha}=0,1,\dots,n-1$. The factor $m^I_{\alpha}$ implies the nodes to which the flavor field couples. We do not have $\mathbb{Z}_n$ symmetry in the presence of the flavor in general. If $N_F^0$ is a multiple of $n$, however, one can distribute the flavors evenly among the different groups and recover $\mathbb{Z}_n$ symmetry.\footnote{For $N_F^0=n$, we can choose $m^I_{\alpha}=\alpha -1$ for $I=1,2$.} Remember that we believe the orbifold equivalence only if $\mathbb{Z}_n$ symmetry is present. Thus, we consider the case in which $N_F^0$ is a multiple of $n$ for the following section.
 We show that the orbifolded theory is given by the $\Nthree$ $[U(N)\times U(N)]^n$ Chern-Simons matter theory with $N_F$ flavors. 
{The orbifold projection of mesonic operators is given by  
\begin{align}\label{MES21}
&\dfrac{1}{pN}\tilde{Q}^1_{\beta}(A_1B_1)^lQ^1_{\beta}\to \dfrac{1}{pN}\sum_{I=odd}{\tilde{Q}}^I_{\alpha}(A_IB_I)^lQ^I_{\alpha}, \quad \\
&\dfrac{1}{pN}\tilde{Q}^1_{\beta}(A_1B_1)^lA_1Q^1_{\beta}\to \dfrac{1}{pN}\sum_{I=odd}{\tilde{Q}}^I_{\alpha}(A_IB_I)^lA_IQ^{I+1}_{\alpha}. \label{MES22}
\end{align}
It can be shown that these operators are in the neutral sector of $\mathbb{Z}_p$ symmetry.}

\section{FREE CHERN-SIMONS MATTER THEORY ON $S^1\times S^2$ }
It is known that the scale, like the critical temperature of the confinement/deconfinement transition, does not change in the large-$N$ orbifold equivalence~\cite{Larsen:2007bm,Unsal:2007fb}. 
In this section, we analyze the orbifold equivalence for the free
Chern-Simons matter theory on $S^1\times S^2$ at finite temperature.
We first analyze the critical temperature of the phase
transition~\cite{S1,A1}. We consider the large-$N$ limit with 't
Hooft coupling $\lambda =N/k \ll 1$. We ignore the contribution of flavor to derive the Hagedorn/deconfinement transition.
 After taking the temporal gauge and integrating out matter, the unitary matrix model
appears from compactifying Chern-Simons matter theory on $S^1\times
S^2$ $(t\sim t+\beta)$,
\begin{align}
&Z=\int \prod^{2p}_{I=1}DU_I\exp \Big[\sum^p_{i=1}\sum^{\infty}_{n=1}\dfrac{1}{n}\Big(z^{unt}(x^n)\mbox{tr}(U_{2i}^n)\mbox{tr}(U_{2i+1}^{-n})+z^{t}(x^n)\mbox{tr}(U_{2i-1}^n)\mbox{tr}(U_{2i}^{-n})  \nonumber \\
&+c.c.\Big)+\dots\Big],
\end{align}
where $\dots$ shows the contribution of flavor and is ignored at present and $t$, $unt$ show single-particle partition
functions for twisted hypermultiplets and untwisted hypermultiplets
as
\begin{align}
z^t_n=z^{unt}_n=z_B(x^n)+(-1)^{n+1}z_F(x^n)\equiv z_n, \quad z_B(x)=\dfrac{2x^{\frac{1}{2}}(1+x)}{(1-x)^2},\quad z_F(x)=\dfrac{4x}{(1-x)^2}, 
\end{align}
where $x=\exp (-\beta)$. 　

The Polyakov loop $U_I=e^{i\beta A_{0,I}}$ satisfies the periodic condition
$U_{2p+1}=U_1$ and $U^{-1}_I=U^{\dagger}_I$. We diagonalize the
eigenvalues of the holonomy matrix $U_I$ as $U_I=\exp (i\theta_{I,a})
$ with $-\pi \le\theta_{I,a} \le \pi$ $(a=1,\dots ,N)$ where in the
large-$N$ limit, each $\theta_I$ is a continuous parameter with a
density $\rho^I(\theta_I)$. The density satisfies
$\int^{\pi}_{-\pi}\rho^I(\theta_I)d\theta_I=1$. Using the density,
the effective action is written as
\begin{align}\label{MAT33}
N^2\sum_{1\le I,J\le 2p}\int d\theta_Id\theta_J'\rho^I(\theta_I)\rho^J (\theta_J)\Big[-\delta_{IJ}\log \Big|\sin \frac{\theta_I-\theta_J'}{2}\Big| +
\sum^{\infty}_{n=1}\dfrac{1}{n}M^{IJ}(z_n)\cos (n(\theta_I-\theta_J')) \Big], 
\end{align}
where $M$ is the $2p\times 2p$ matrix $M$ as
\begin{align}
M=\begin{pmatrix} 0 & -z_n & 0 & \dots & -z_n \\
-z_n & 0 & -z_n & \dots & 0 \\
 0 & -z_n & 0 & \dots & 0 \\ 
  & \vdots &  & \ddots &                 \\
 -z_n & 0 & 0 & \dots & 0 \end{pmatrix}. 
\end{align}
The first term in Eq.~\eqref{MAT33} is obtained from the change of the measure. 

Using the Fourier transformation
$\rho^I_n=\int d\theta_I \rho^I(\theta_I)\cos(n\theta_I)$, 
the effective action is rewritten as
\begin{align}
&S_{eff}=\sum_n\dfrac{N^2}{n}\rho^I_n(\delta^{IJ}+ M^{IJ}) \bar{\rho}^J_{n},
\end{align}
 and 
 $(\delta +M)_{IJ}$ satisfies a circulant determinant formula
\begin{align}
&\det\begin{pmatrix} g_1 & g_2 & g_3 & \dots & g_{k} \\
g_{k} & g_1 & g_2 & \dots & g_{k-1} \\
 g_{k-1} & g_k & g_1 & \dots & g_{k-2} \\ 
  & \vdots &  & \ddots &                 \\
 g_2 & g_3 & g_4 & \dots & g_1 \end{pmatrix}=\prod_{I=0}^{k-1}(g_1+\omega^Ig_2+\omega^{2I}g_3+\dots +\omega^{(k-1)I}g_k), \\
& \det (1+M)=(1-2z_n)\prod^{2p-1}_{I=1}(1-z_n\omega^I-z_n\omega^{(2p-1)I}),
\end{align}
where $\omega=\exp (\pi i/p)$.

The phase structure can be analyzed using the above effective action. 
At low temperature, the real eigenvalues of $M$ are positive, and
so the trivial saddle point $\rho_n =0$ is dominated when $2z_1<1$
as $z_n$ is a monotonically decreasing function of $n$. The 
$O(N^2)$ contributions are not included in this saddle point, and
the Casimir energy vanishes for the above $d=3$ Chern-Simons matter
theory. For $2z_1>1$, one of the eigenvalues becomes negative, and
another saddle point  where the free energy is of order $O(N^2)$
is dominated. The confinement/deconfinement phase transition happens
at $2z_1=1$. Note that the phase transition happens at the same
critical temperature $T_H=1/\log (17+12\sqrt{2})$ as that of the
free ABJM theory analyzed in Ref.~\cite{NTP}. This implies that $\mathbb{Z}_p$
orbifold symmetry is not broken in the free theory on $S^1\times
S^2$. We show this unbroken
$\mathbb{Z}_p$ symmetry slightly above critical temperature and at high temperature from now on. 
 
Slightly above the critical temperature, the density is nonzero
only for $-\theta_{Ic}\le \theta_I \le \theta_{Ic}$. The saddle
point of $\theta_I$ is obtained by solving the following equation:
\begin{align}\label{SAE38}
\int d\theta'_I\rho ^I (\theta_I')\cot \Big(\dfrac{\theta_I-\theta_I'}{2}\Big)=-2
\sum _J\sum^{\infty}_{n=1}M_{IJ}\sin (n\theta_I)\rho^J_n.
\end{align}
To obtain the free energy, the approximation $z_n=0$ for $n>1$ is
used. Namely, only the first winding state in the time direction
is excited, and this approximation is valid at temperatures that are not high.
 Reflecting the $\mathbb{Z}_p$ symmetry of the Chern-Simons matter theory,
moreover, the densities of eigenvalues should take the same form, 
$\rho^I(\theta)=\rho (\theta)$.
Using the above approximation and $\mathbb{Z}_p$ symmetry, Eq.~\eqref{SAE38} is rewritten as
\begin{align}\label{SAE39}
\int d\theta'_I\rho (\theta_I')\cot \Big(\dfrac{\theta_I-\theta_I'}{2}\Big)=4
z_1\sin (\theta_I)\rho_1.
\end{align}
The solution for Eq.~\eqref{SAE39} is given by
\begin{align}\label{DEN310}
\rho (\theta)=\dfrac{1}{\pi \sin^2\dfrac{\theta_c}{2}}\sqrt{\sin^2\dfrac{\theta_c}{2}-\sin^2\dfrac{\theta}{2}}\cos\dfrac{\theta}{2}
\end{align}
for $-\theta_{Ic}\le \theta_I \le \theta_{Ic}$. Here, $\theta_c$ satisfies
\begin{align}
\sin^2\dfrac{\theta_c}{2}=1-\sqrt{1-\dfrac{1}{2z_1}},
\end{align}
where the quantity inside the root is positive for $2z_1>1$. Using the $\mathbb{Z}_p$
symmetry of the densities of eigenvalues, the free energy $F=-\log
Z/\beta$ is given by
\begin{align}
 F=-\dfrac{2pN^2}{\beta}\Big(\dfrac{1}{2\sin^2\frac{\theta_c}{2}}+\dfrac{1}{2}\log \Big(\sin^2\dfrac{\theta_c}{2}\Big)-\dfrac{1}{2}\Big).
\end{align}
From the above formula, we can show the relation expected in the
orbifold equivalence between the $U(pN)\times U(pN)$ ABJM theory
and the $[U(N)\times U(N)]^p$ Chern-Simons matter theory as
\begin{align}
F=\dfrac{F^{ABJM}}{p},
\end{align}
where $F^{ABJM}$ is the free energy of the $U(pN)\times U(pN)$ ABJM
theory. We also compare the Polyakov loop VEV between the
Chern-Simons matter theories which are equivalent under the orbifold
equivalence. The Polyakov loop VEV normalized by the rank is given by
the first moment $\rho_1$ of $\rho (\theta)$~\cite{Gross:1980he}.
It can be obtained from Eq.~\eqref{DEN310} as
\begin{align}
\dfrac{\langle \mbox{tr}(U_I) \rangle}{N}=\rho_1=\begin{cases} & \dfrac{1}{4z_1(1-\sqrt{1-\frac{1}{2z_1}})}\quad 2z_1\ge 1 \\
& 0\quad  2z_1 \le 1
\end{cases}.
\end{align}
Remember that $\rho_1$ becomes $1/2$ at the critical point $2z_1=1$.
Since $\rho_1$ does not depend on the orbifold action, the above result is consistent with the orbifold equivalence as
\begin{align}
\langle \mbox{tr}(U_I) \rangle =\dfrac{\langle \mbox{tr}(U_I^{ABJM}) \rangle }{p}.
\end{align}

{Finally, in the high-temperature limit, the free energy of the Chern-Simons matter theory is obtained using the different saddle point $\rho_n =1$ for all $n$. The result agrees with the relation expected in the orbifold equivalence as
\begin{align}
F=-28pT^3\zeta (3)N^2,\quad F=\dfrac{F^{ABJM}_2}{p},
\end{align}
where $F^{ABJM}_2$ is the free energy of the ABJM theory. 

We can also include flavor for the analysis of the above free theory, which changes the order of the confinement/deconfinement transition~\cite{Schnitzer:2004qt,Basu:2008uc,Liu:2004vy}. When including flavor, it is known that large-$N$ orbifold equivalence still  holds if $\mathbb{Z}_p$ symmetry is unbroken.} 

\section{THE GRAVITY DUAL TO THE $\Nfour$ THEORY}
According to Ref.~\cite{Imamura:2008ji}, we verify the action of the above orbifold in the dual gravity side and show the orbifold equivalence between the ABJM theory and $\Nfour$ Chern-Simons matter theory with the equal amount of twisted and untwisted hypermultiplets introduced in Sec. II via holography.
 We consider the dual $AdS_4\times S^7/\mathbb{Z}_{pk}\times \mathbb{Z}_p$ geometry of $\Nfour$ SCFT constructed via type-IIB $N$ D3, $p$ NS5, and $p$ $(1,k)$ 5-branes.
This SCFT corresponds to $\Nfour$ Chern-Simons matter theory with the equal amount of twisted and untwisted hypermultiplets. 
  It is convenient to represent $S^7/\mathbb{Z}_{pk}\times \mathbb{Z}_p$ in terms of four complex coordinates $X_i$ $(i=1,2,3,4)$ as
 \begin{align}
 &X_1=\cos\xi\cos \dfrac{\theta_1}{2}e^{i\frac{\chi_1+\varphi_1}{2}},\quad X_2=\cos\xi \sin\dfrac{\theta_1}{2}e^{i\frac{\chi_1 -\varphi_1}{2}}, \quad \\
 &X_3=\sin\xi\cos \dfrac{\theta_2}{2}e^{i\frac{\chi_2+\varphi_2}{2}},\quad X_4=\sin\xi\sin \dfrac{\theta_2}{2}e^{i\frac{\chi_2-\varphi_2}{2}},
 \end{align}
 where  $0\le \xi<\pi/2$, $(\chi_1,\chi_2)\sim (\chi_1+\frac{4\pi}{kp},\chi_2 +\frac{4\pi}{kp})\sim (\chi_1+\frac{4\pi}{p},\chi_2 )$, $0\le \varphi_i<2\pi$, and $0\le \theta_i <\pi$. 
The $\mathbb{Z}_{kp}$ and $\mathbb{Z}_p$ orbifold action is written in terms of $X_i$ as
\begin{align}
&(X_1,X_2,X_3,X_4)\sim e^{2\pi i/(kp)}(X_1,X_2,X_3,X_4),\\ 
&(X_1,X_2,X_3,X_4)\sim (e^{2\pi i/p}X_1,e^{2\pi i/p}X_2,X_3,X_4).
\end{align}
According to Ref.~\cite{Hikida:2009tp}, $X_i$ can be identified with the complex parameters $y^A$ parametrizing $\mathbb{C}^4$. Using this identification, the isometry corresponding to $SU(2)\times SU(2)$ $R$ symmetry of the $\Nfour$ SCFT rotates $(X_1,X_2)$ and $(X_3,X_4)$, respectively. The orbifold actions Eqs.~\eqref{ORB22} and \eqref{ORB23} for $y^A$ are also consistent with the above identification.
 
After considering the backreaction of $N$ M2 branes and the near-horizon limit, 
 the dual geometry of $\Nfour$ Chern-Simons matter theory is described by
\begin{align}
 &ds^2_{11D}=\dfrac{R^2}{4}ds^{2}_{AdS_4}+R^2ds^2_7,\quad R=l_p(2^5Nkp^2\pi^2)^{1/6}, \nonumber \\
 &ds^2_7=d\xi^2 +\dfrac{1}{4}\cos^2\xi ((d\chi_1+\cos\theta_1d\varphi_1)^2+d\theta_1^2+\sin^2\theta_1d\varphi_1^2) \nonumber \\
 &+\dfrac{1}{4}\sin^2\xi ((d\chi_2+\cos\theta_2d\varphi_2)^2+d\theta_2^2+\sin^2\theta_2d\varphi_2^2).
 \end{align}
The isometry of $ds^2_7$ is $(SU(2)\times U(1))^2$ and is interpreted as the global symmetry of $\Nfour$ SCFT. To make the classical M-theory description valid, the size of the orbifolded M circle in terms of $\mathbb{Z}_{pk}$ must be larger than the 11-dimensional Planck length, $R/(l_pkp)\gg 1$ or $pN/(kp)^5\gg 1$.
 
According to Ref.~\cite{Fujita:2009xz}, we also consider the Kaluza-Klein reduction to type-IIA string theory, setting $\alpha' =1$ as
\begin{align}
&ds^2_7=ds^2_6+\dfrac{1}{k^2p^2}(dy+A)^2,\quad e^{2\phi}=\dfrac{R^3}{k^3p^3}, \nonumber \\
&A=kp \Big( \dfrac{1}{2}\cos^2\xi (d\psi +\cos\theta_1 d\varphi_1)+\dfrac{1}{2}\sin^2\xi \cos\theta_2 d\varphi_2 \Big), \nonumber \\
&ds^2_6=d\xi^2 +\dfrac{1}{4}\cos^2 \xi \sin^2\xi (d\psi +\cos\theta_1 d\varphi_1 -\cos\theta _2 d\varphi_2)^2 \nonumber \\
&+\dfrac{1}{4}\cos^2\xi (d\theta_1^2+\sin^2\theta_1d\varphi_1^2)+\dfrac{1}{4}\sin^2\xi (d\theta_2^2+\sin^2\theta_2d\varphi_2^2), \nonumber \\
&ds^2_{IIA}=L^2(ds^2_{AdS_4}+4ds^2_6),\quad L^2=\dfrac{R^3}{4kp}=2^{1/2}\pi \sqrt{\dfrac{N}{k}}, \nonumber \\
&0< {y}\le {2\pi},\quad 0<\psi \le \dfrac{4\pi}{p},\quad \chi_1
=\psi +\dfrac{2y}{kp},\quad \chi_2 =\dfrac{2y}{kp}.
\end{align}
Remember that we do not have the orbifold fixed point in terms of the $\mathbb{Z}_{kp}$ action like with the ABJM theory, since the size of the orbifolded M circle is constant. 
The weakly coupled type-IIA theory description is valid when the size of the orbifolded M circle or the string coupling is small, $pN/(kp)^5\ll 1$.
Remember that the metric $ds^2_6$ is equal to the orbifold $\mathbb{CP}^3/\mathbb{Z}_p$. The D2-brane flux and the D6-brane flux $(F_2=dA)$ are quantized as $N\ (=\int* F_4/(2\pi)^5)$ and $kp\ (=\int F_2/(2\pi))$, respectively, consistent with the brane configuration since we must have $p$ $(1,k)$ 5-branes in the type-IIB elliptic D3-brane configuration through $T$ duality. The curvature radius $2^{1/2}\pi\sqrt{N/k}$ of $AdS_4\times \mathbb{CP}^3/\mathbb{Z}_p$ in the type-IIA description should be large to make the supergravity description valid. That is, the type-IIA description is valid when $pk \ll pN \ll (pk)^5$.

Hereby, we discuss the orbifold equivalence between the $\Nfour$ $[U(N)_{k}\times U(N)_{-k}]^p$ Chern-Simons matter theory and the $U(pN)_{pk}\times U(pN)_{-pk}$ ABJM theory. 
 Note that the curvature radius of the above type-IIA solution is equal to the curvature radius of the type-IIA solution $2^{1/2}\pi\sqrt{N/k}$ dual to the $U(pN)_{pk}\times U(pN)_{-pk}$ ABJM theory. In addition, the D6-brane flux of the latter theory is the same as that of the former theory $pk$.  
  Thus, the type-IIA geometry for $\Nfour$ SCFT is related with the type-IIA geometry of the ABJM theory by the $\mathbb{Z}_p$ orbifold.\footnote{Actually, the region in which both the type-IIA theory and the M-theory descriptions are valid is precisely the same form.} The equivalence should work for any observables that are invariant under the $\mathbb{Z}_p$ projection.
    
{When we include probe D6 branes dual to flavor, we also need to take care of $\mathbb{Z}_p$ discrete symmetry in the presence of flavor. For $\Nfour$ SCFT with $2p$ nodes, the D6 brane corresponding to the massless flavor wraps $AdS_4\times S^3/\mathbb{Z}_{2p}$ inside $AdS_4\times \mathbb{CP}^3/\mathbb{Z}_{p}$ without a  tadpole problem~\cite{Fujita:2009xz} (see also Refs.~\cite{Gaiotto:2009tk,Hikida:2009tp}). In large-$N$ orbifold equivalence between the ABJM theory and $\mathcal{N}=4$ SCFT with $2p$ nodes, we should add $2pN_F$ D6 branes to recover $\mathbb{Z}_p$ discrete symmetry of the orbifold. Moreover, we find that the node coupling to fundamentals can be specified using the holonomy of the Wilson line\footnote{We would like to thank A. Karch for pointing out this point.} $\pi_1 (S^3/\mathbb{Z}_{2p})=\mathbb{Z}_{2p}$. We can show that in large-$N$ orbifold equivalence, the fluctuations of the D6 brane in the neutral sector in the presence of the $\mathbb{Z}_{2p}$ holonomy coincide between the parent theory and the daughter theory. The mesonic operators Eqs.~\eqref{MES21} and \eqref{MES22} correspond to the operators dual to the scalar fluctuations where Eq.~\eqref{MES22} corresponds to the mode including the $\mathbb{Z}_{2p}$ holonomy. 
 As an application, the holographic BKT phase transition~\cite{Jensen:2010vx} in terms of the ABJM with flavor can be applied in large-$N$ orbifold equivalence, since the fluctuation around the massless embedding corresponds to the neutral sector of the orbifold. We leave the general analysis, including massive flavor where the submanifold wrapped by the D6 brane is squashed in terms of the mass parameter, for future work. We can use the same argument of the holonomy when this submanifold is topologically equivalent to $S^3/\mathbb{Z}_{2p}$. }

\section{THE GRAVITY DUAL TO THE $\Nthree$ CHERN-SIMONS MATTER THEORY WITH FLAVOR}
In this section, we review the $d=8$ transverse geometry of M2 branes
describing the $\Nthree$ quiver Chern-Simons matter theory with flavor. This transverse
geometry becomes the $d=8$ toric hyperK{\"a}hler manifold, where toric
means there is at least a two-torus inside it.
Using the cone structure of this transverse space, we also observe
that the Eschenburg space which is tri-Sasaki manifold gives the
gravity dual of the $\Nthree$ quiver Chern-Simons matter theory with flavor.

The metric of the $d=8$ toric hyperK{\"a}hler manifold ($\varphi_i\in
(0,4\pi]$ ) is given by
\begin{eqnarray}\label{MM}
\begin{cases}
ds^2 = \frac{1}{2} U_{ij} 
d {\boldsymbol{x}}_i \cdot  d {\boldsymbol{x}}_j + \frac{1}{2}U^{ij} (d\varphi_i + A_i)(d\varphi_j +A_j )\\
A_i=d{\boldsymbol{x}}_j \cdot \boldsymbol{\omega}_{ji}
=dx^a_j ~ \omega_{ji}^a, \quad \partial_{x^a_j}\omega^b_{ki}-\partial_{x^b_k} \omega_{ji}^a=\epsilon^{abc}\partial_{x^c_j} U_{ki},
\end{cases}
\end{eqnarray}
where $i,j,k=1,2$, $a,b,c=1,2,3$, and $U^{ij}$ is the inverse matrix of $U_{ij}$.

If we introduce $p$ NS5 and $p$ $(1,k)$5-branes in a type-IIB brane setup, we have
\begin{eqnarray}
U_{ij}=\frac{1}{2}\left(
\begin{array}{cc}
\dfrac{p}{|{\boldsymbol{x}}_1| } +  
\dfrac{p}{|{\boldsymbol{x}}_1 + k{\boldsymbol{x}}_2|} & 
\dfrac{kp}{|{\boldsymbol{x}}_1 + k{\boldsymbol{x}}_2|}\\
\dfrac{kp}{|{\boldsymbol{x}}_1 + k{\boldsymbol{x}}_2|}&
\dfrac{k^2 p}{|{\boldsymbol{x}}_1 + k{\boldsymbol{x}}_2|}
\end{array}
\right),
\label{u}
\end{eqnarray}
where $U$ is normalized, being consistent with the quantization
condition on the flux of $\boldsymbol{\omega}$. 
We can use the $GL(2)$ transformation to diagonalize the matrix $U$.
\begin{eqnarray}
\label{UDG}
&({\boldsymbol{x}}'_1, {\boldsymbol{x}}'_2) = 
({\boldsymbol{x}}_1, {\boldsymbol{x}}_2) G^t=p({\boldsymbol{x}}_1, {\boldsymbol{x}}_1+k{\boldsymbol{x}}_2) , \nonumber  \\
&({{\varphi}}'_1, {{\varphi}}'_2) = ({{\varphi}}_1, {{\varphi}}_2) G^{-1}=\Big(\dfrac{{{\varphi}}_1}{p}-\dfrac{{{\varphi}}_2}{kp}, \dfrac{{{\varphi}}_2}{kp}\Big) ,\nonumber \\
&G=\left(
\begin{array}{cc}
p & 0 \\
p & kp 
\end{array}
\right),  \quad
U ~\to ~ U'= \frac{1}{2}\left(
\begin{array}{cc}
\dfrac{1}{|{\boldsymbol{x}}'_1|}&0\\
0&\dfrac{1}{|{\boldsymbol{x}}'_2|}
\end{array}
\right). 
\end{eqnarray}
The orbifold action of $(\varphi_1',\varphi_2')$ is given by 
\begin{align}\label{ORB16} (\varphi_1',\varphi_2')\sim \Big(\varphi_1'+\dfrac{4\pi}{p},\varphi_2'\Big),\quad  (\varphi_1',\varphi_2')\sim \Big(\varphi_1'-\dfrac{4\pi}{kp},\varphi_2'+\dfrac{4\pi}{kp}\Big) \end{align}

The contribution of $N_F$ flavor is included by adding the following
extra $\Delta U$ to $U$:
\begin{align}
    \Delta U
    =\frac{1}{2}\left(
    \begin{array}{cc}
	0&0\\
	0&\dfrac{N_F }{|{\boldsymbol{x}}_2 |}
    \end{array} 
    \right).
\end{align}  

Here, the $GL(2)$ transformation Eq.~\eqref{UDG} acts on $\Delta U$ as
\begin{eqnarray*}
\Delta U ~\to ~\Delta U'
=\frac{1}{2}\left(
\begin{array}{cc}
\dfrac{ N_F }{kp\SL}
&\dfrac{- N_F }{kp\SL}\\
\dfrac{- N_F }{kp\SL}
&\dfrac{ N_F }{kp\SL}
\end{array}
\right),  ~~~~~~~~~~~~
\SL=|{\boldsymbol{x}}'_2 - {\boldsymbol{x}}'_1|.
\end{eqnarray*}
Here, we see the following symmetry in the above metric. Since $N_F$
is nonzero, there is a common element of $SO(3)$ which rotates
$({\boldsymbol{x}}'_1, {\boldsymbol{x}}'_2)$ in order to preserve
$\SL$. Furthermore, we generate $U(1)_b \times U(1)_d$ corresponding
to two $U(1)$'s of $(\varphi'_1 ,\varphi'_2 )$ by using 
gauge transformations of
$(A'_1, A'_2)$: $(A'_1, A'_2)=(A'_1+d\lambda_1, A'_2+d\lambda_2)$. 
It is convenient to use the transformation
${\boldsymbol{x}}'_1\to -{\boldsymbol{x}}'_1$,
${{\varphi}}'_1 \to -{{\varphi}}'_1$
and perform the following reparametrization in terms of three $t$'s:
\begin{eqnarray*}
\Delta U'= \frac{1}{2} \left(
\begin{array}{cc}
\dfrac{t_1^2 }{t_3 |t_1 {\boldsymbol{x}}'_1 + t_2 {\boldsymbol{x}}'_2| }
&\dfrac{t_1 t_2  }{t_3 |t_1 {\boldsymbol{x}}'_1 + t_2 {\boldsymbol{x}}'_2| }\\
\dfrac{t_1 t_2  }{t_3 |t_1 {\boldsymbol{x}}'_1 + t_2 {\boldsymbol{x}}'_2| }
&\dfrac{t_2^2}{t_3 |t_1 {\boldsymbol{x}}'_1 + t_2 {\boldsymbol{x}}'_2| }
\end{array}
\right),
\end{eqnarray*}
where $(t_1, t_2, t_3)=(N_F,N_F,kp)$ are relatively prime without a divisor as seen in the  above metric components~\cite{Fujita:2009xz}. Using $V=U'+\Delta U'$, 
the metric in Eq.~\eqref{MM} is rewritten as
\begin{align}\label{MM2}
ds^2 = \frac{1}{2} V_{ij} 
d {\boldsymbol{x}'}_i \cdot  d {\boldsymbol{x}'}_j + \frac{1}{2}V^{ij} (d\varphi_i' + A_i')(d\varphi_j' +A_j' ).
\end{align}
It can be shown that  
$ds^2$ in Eq.~\eqref{MM2} becomes the cone over a $d=7$ Eschenburg space 
$\mathcal{M}_7=S_7^{(p,kp)}(N_F,N_F,kp)$ with the orbifold action Eq.~\eqref{ORB16},\footnote{In the previous paper~\cite{Fujita:2009xz}
we did not clarify the inclusion of the orbifold. However, this
orbifold does not change the volume of the Eschenburg space.}
namely, $\mathbb{Z}_{p}\times \mathbb{Z}_{kp}$. Note that unlike the ABJM theory, we have the fixed point at the $\mathbb{Z}_k$ singurality~\cite{Cheon:2011th} when $k\neq 1$ (see also the Appendix of Ref.~\cite{Fujita:2009xz}), and it leads to the light degrees of freedom on the fixed points beyond the supergravity approximation. However, we do not consider these light modes in this paper, since the large-$N$ orbifold equivalence is applied for only the untwisted sector. $ds^2$ in Eq.~\eqref{MM2}
has the isometry $SO(3) \times SU(2)_d\times U(1)_b$, where we have
$SU(2)_d$ enhanced global symmetry instead of $U(1)_d$~\cite{Lee:2006ys}, 
since $t_1=t_2\neq t_3$.

Note that by replacing $kp$ with $k'$ in the metric Eq.~\eqref{MM2},
we have the cone over an  Eschenburg space
$\mathcal{M}_7=S_7^{(p,k')}(N_F,N_F,k')$. Note that this $\mathcal{M}_7$
is the $\mathbb{Z}_p$ orbifold of the $p=1$ case,
$\mathcal{M}_7^{(p=1)}=S_7^{(1,k')}(N_F,N_F,k')$. In other words,
operating the $\mathbb{Z}_{p}$ orbifold on $\mathcal{M}_7^{(p=1)}$,
both the metrics become the same. It implies that the $\mathbb{Z}_p$
orbifold equivalence works as seen in the next section.

We can also consider the replacement of $N_F$ with $N_F'k$. Then,
we have the cone over an Eschenburg space
$\mathcal{M}_7=S_7^{(p,kp)}(N_F',N_F',p)$. Note that we have different
charges because three charges $(N_F'k,N_F'k,pk)$ are not relatively
prime. This $\mathcal{M}_7$ is exactly the $\mathbb{Z}_{kp}$ orbifold
of the $k=1$ case, $\mathcal{M}_7^{(k=1)}=S_7^{(p,p)}(N_F',N_F',p)$
instead of the $\mathbb{Z}_p$ orbifold.

After including the backreaction of $N$ M2 branes on the $d=8$
transverse space and taking the near-horizon limit, we obtain
$AdS_4\times {\mathcal M}_7$,
\begin{eqnarray}
&&ds^2_{11D} = \frac{R^2}{4} ds^2_{AdS_4} + R^2 ds_7^2 , \quad\quad N=\frac{1}{(2\pi \ell_p)^6}\int_{{\mathcal M}_7} \ast F_4, \nonumber \\
&& F_4=\frac{3}{8} R^3 vol_{AdS_4},\quad R^6 vol({\mathcal M}_7)=(2\pi \ell_p)^6 N,
\label{NOR}
\end{eqnarray}
where the relation $R_{ab} ({\mathcal M}_7) =6g_{ab} ({\mathcal M}_7)$ is satisfied and $R=2R_{AdS}$ is the radius of ${\mathcal M}_7$. This background is the gravity dual of the strongly coupled limit of the ${\mathcal N}=3$ Chern-Simons matter theory. 

\subsection{The $\mathbb{Z}_p$ orbifold equivalence}

According to Ref.~\cite{Lee:2006ys}, we have the following relation of the volume of the Eschenburg space and the radius $R$:
\begin{eqnarray}
\frac{{{vol}}({S^7})}{vol({{{\mathcal M}}_7})}
=\frac{2p(N_F+kp)^2}{\big( N_F + 2kp\big)},\quad R^6=2^5\pi^2pNl_p^6 \cdot\frac{2(N_F+kp)^2}{\big( N_F + 2kp\big)},
\label{vol}
\end{eqnarray}
where ${vol}({S^7})=\pi^4/3$. 
It is interesting to consider the case of $p=1$, namely, $\mathcal{M}_7^{(p=1)}=S_7^{(1,k')}(N_F,N_F,k')$. 
 The volume and the radius $R'$ of $\mathcal{M}_7^{(p=1)}$ are given by
\begin{eqnarray}
\frac{{{vol}}({S^7})}{vol({{{\mathcal M}}_7^{(p=1)}})}
=\frac{2(N_F+k')^2}{\big( N_F + 2k'\big)},\quad R^{\prime 6}=2^5\pi^2Nl_p^6 \cdot\frac{2(N_F+k')^2}{\big( N_F + 2k'\big)}.
\label{vol2}
\end{eqnarray}  
Note that by replacing $kp$ with $k'$ in the metric Eqs.~\eqref{MM2} and
\eqref{vol}, the volume of the Eschenburg space
$vol(\mathcal{M}_7^{(p=1)})$ is found to be the product of
$vol({\mathcal M}_7)$ and the orbifold factor $p$ as
\begin{eqnarray}
vol(\mathcal{M}_7)p=vol(\mathcal{M}_7^{(p=1)}).  \label{vol11}
\end{eqnarray} 
Using Eq.~\eqref{vol11}, we can show that the radius $R$ of $AdS_4\times
\mathcal{M}_7$ is equal to the radius of $AdS_4\times
\mathcal{M}_7^{(p=1)}$, setting $pN$ M2-brane flux of the latter
theory. Thus, the formula Eq.~\eqref{vol2} shows that the M2 brane
theory with $N$ M2 branes and $\mathcal{M}_7=S_7^{(p,k')}(N_F,N_F,k')$
are equivalent to that with $pN$ M2 branes and
$\mathcal{M}_7^{(p=1)}=S_7^{(1,k')}(N_F,N_F,k')$ in terms of the
orbifold equivalence, since the metric of the Eschenburg space becomes
the same up to the $\mathbb{Z}_p$ orbifold.
Thus, we observe the structure of the orbifold equivalence between
the $\Nthree$ $U(pN)_{k'}\times U(pN)_{-k'}$ quiver Chern-Simons matter theory with
$N_F$ flavor and the $\Nthree$ $(U(N)_k\times U(N)_{-k})^p$ quiver Chern-Simons matter
theory with $N_F$ flavor.

\subsection{The $\mathbb{Z}_{kp}$ orbifold equivalence}
The case $k=1$ is also interesting, namely, $\mathcal{M}_7^{(k=1)}=S_7^{(p,p)}(N_F',N_F',p)$.  
 The volume and the radius $R'$ of $\mathcal{M}_7^{(k=1)}$ are obtained from Eq.~\eqref{vol} as
\begin{eqnarray}
\frac{{{vol}}({S^7})}{vol({{{\mathcal M}}_7^{(k=1)}})}
=\frac{2p(N_F'+p)^2}{\big( N_F' + 2p\big)},\quad R^{\prime 6}=2^5\pi^2pNl_p^6 \cdot\frac{2(N_F'+p)^2}{\big( N_F' + 2p\big)}.
\label{vol110}
\end{eqnarray}  
Note that by substituting $N_F=kN_F'$ into the metric Eqs.~\eqref{MM2} and \eqref{vol}, the volume of the Eschenburg space $vol(\mathcal{M}_7^{(k=1)})$ becomes the product of $vol({\mathcal M}_7)$ and the orbifold factor $k$ as
\begin{eqnarray}
vol(\mathcal{M}_7)k=vol(\mathcal{M}_7^{(k=1)}).
\label{vol112}
\end{eqnarray}
Using Eq.~\eqref{vol112}, it can be shown that the curvature radius $R$ of $AdS_4\times \mathcal{M}_7$ is equal to the curvature radius of $AdS_4\times \mathcal{M}_7^{(k=1)}$, setting $kN$ M2-brane flux of the latter theory.  Thus, the formula Eq.~\eqref{vol110} implies the orbifold equivalence between $N$ M2 branes with $S_7^{(p,kp)}(N_F,N_F,kp)$ and $kN$ M2 branes with $S_7^{(p,p)}(N_F',N_F',p)$,  since the metric of the Eschenburg space becomes the same up to the $\mathbb{Z}_{kp}$ orbifold. 
 Thus, we find that the $\Nthree$ $(U(kN)_{1}\times U(kN)_{-1})^{p}$ quiver Chern-Simons matter theory with $N_F'$ flavor is equivalent to the $\Nthree$ $(U(N)_{k}\times U(N)_{-k})^p$ quiver Chern-Simons matter theory with $N_F=N_F'k$ flavor in view of the orbifold equivalence. 
   
As an example, we consider the BPS observables of Chern-Simons matter theories including flavor invariant under the $\mathbb{Z}_{k}$ orbifold projection. We start with $[U(N)_k\times U(N)_{-k}]^p$ and $kN_F'$ fundamentals. There are operators charged under $U(1)_b$ and operators neutral under $U(1)_b$. Hereby, we concentrate on the operators charged under $U(1)_b$. The operator dual to the D0 brane is the operator with smallest dimension and charged under $U(1)_b$~\cite{Gaiotto:2009tk}.

Note that in the Abelian case, the Chern-Simons EOM is satisfied via a constant magnetic flux $m$ on the sphere. 
Thus, we can construct the operators charged under $U(1)_b$ by introducing diagonal monopole operators. Such a monopole operator is shown to be BPS and is defined as $T^{(m)}$ with the same monopole flux $m (\in \mathbb{Z}_n)$ under all $U(1)^{2p}$ subgroups of $U(N)^{2p}$ gauge groups~\cite{Benna:2009xd}. In the Chern-Simons matter theories with $\sum k_i=0$, the monopole $T^{(m)}$ has the charges $(mk,-mk,\dots ,-mk)$ under the $U(1)^{2p}$ subgroup. Moreover, it is known that the monopole operators $T^{(m)}$ can be charged under any $U(1)$ symmetry via quantum corrections~\cite{Borokhov:2002cg,Borokhov:2003yu}. 
 In the Abelian case, the quantum correction to the R charge of the monopole operators is~\cite{Benini:2009qs}
\begin{align}
\delta R[T^{(m)}] =-\dfrac{m}{2}\sum_{\psi}R[\psi]=\dfrac{mkN_F'}{2},
\end{align}
where we have summed over the R charge for all fermions. We use the fact that the gaugino has the charge 1, which is canceled by the R charges of bifundamentals. 

The gauge-invariant operators are of the form 
\begin{align}\label{MON51}
T^{(m)}\prod_i A_i^{d_i} \quad \text{for $d_{i}-d_{i+1} =k_im$},
\end{align}
where $d_i$ is proportional to the D5 brane charges of the $i$ th 5-branes. The  above operator has the baryonic charge $\sum_id_i= mkp$.
 When $m=1$, the operator Eq.~\eqref{MON51} is dual to the D0 brane. The conformal dimension of the operators in Eq.~\eqref{MON51} is given by 
\begin{align}
\Delta=\dfrac{1}{2}(\sum d_i +mkN_F')=\dfrac{mk}{2}(p+N_F').
\end{align}
Namely, the operators preserving the number $m_1k_1=m_2k_2$ under the orbifold action are the invariant operators under the $\mathbb{Z}_{k}$ orbifold projection which does not change the number of nodes. Especially, the operator with flux $m>1$ in the parent theory can be mapped onto the operator dual to the D0 brane in the daughter theory.

\subsection{Orbifold equivalence in terms of entropy}

In this subsection, we show that the orbifold equivalence works for
the Bekenstein-Hawking entropy.
We consider the $AdS$-Schwarzschild black hole metric as
 \begin{align}
 ds^2=\Big(\dfrac{4r^2}{R^2}+1-\dfrac{M}{r}\Big)d\tau^2 +\dfrac{dr^2}{\Big(\dfrac{4r^2}{R^2}+1-\dfrac{M}{r}\Big)}+r^2d\Omega_2^2. \label{MET45}
 \end{align}
The above metric describes the finite-temperature Chern-Simons matter theory on $S^1\times S^2$. The inverse temperature $\beta$ is given by 
\begin{align}
\beta =\dfrac{\pi R^2r_0}{3r_0^2+\frac{R^2}{4}}, \label{TEM46}
\end{align}
where $r_0$ describes the horizon radius. Solving Eq.~\eqref{TEM46}, the horizon radius is represented as
\begin{align}
r_0=\dfrac{\pi R^2}{6\beta}+\sqrt{\Big(\dfrac{\pi R^2}{6\beta}\Big)^2-\dfrac{R^2}{12}}. \label{RAD47}
\end{align}
From Eq.~\eqref{RAD47}, we find that $AdS$ black holes exist at the temperatures larger than $\sqrt{3}/(\pi R)$. 

Using the metric in Eq.~\eqref{NOR}, the Bekenstein-Hawking area law gives
the entropy of $N$ M2 branes on the singularity of Cone$(\mathcal{M}_7)$ per $vol
(S^2)R^2/4$ as (see also Ref.~\cite{Gubser:1996de})
\begin{align}
S_{E}=\dfrac{2^{\frac{3}{2}}\pi^4N^{\frac{3}{2}}}{3^{\frac{7}{2}}\beta^2\sqrt{vol (\mathcal{M}_7)}}\Big(1+\sqrt{1-\dfrac{3\beta^2}{\pi^2R^2}}\Big). \label{ENT417}
\end{align} 
  
We compare Eq.~\eqref{ENT417} with the cases for the M theory on
$AdS_4\times \mathcal{M}_7^{(k=1)}(\mathcal{M}_7^{(p=1)})$, where
we have $pN$($kN$) numbers of the M2 brane flux. Defining the
corresponding entropy
 $S^{(p=1)}_E$($S^{(k=1)}_E$) for each case, we obtain the following relations:
\begin{align}
\dfrac{S_E}{S^{(p=1)}_E}=\dfrac{1}{p}\sqrt{\dfrac{vol(\mathcal{M}_7^{(p=1)})}{p \cdot vol(\mathcal{M}_7)}}=\dfrac{1}{p},\quad \dfrac{S_E}{S^{(k=1)}_E}=\dfrac{1}{k}\sqrt{\dfrac{vol(\mathcal{M}_7^{(k=1)})}{k\cdot vol(\mathcal{M}_7)}}=\dfrac{1}{k}, \label{ENT41}
\end{align}
where we used the relations in Eqs.~\eqref{vol11} and \eqref{vol112} among
$vol(\mathcal{M}_7)$, $vol(\mathcal{M}_7^{(k=1)})$, and
$vol(\mathcal{M}_7^{(p=1)})$. Usually, the orbifold just affects
the geometry through a projection in the internal space, which
changes its volume, and this will be reflected in the entropy. Thus,
the above relations are those expected in the orbifold equivalence. {Note that the entropy Eq.~\eqref{ENT417} in the M-theory region is consistent with the planar equivalence outside the planar limit~\cite{Azeyanagi:2012xj}. It implies that large-$N$ equivalence holds even outside the 't Hooft limit as long as there is a classical gravity dual.}

We can see that the orbifold equivalence also works until the
Hawking-Page transition happens.
It is known that the Hawking-Page
transition~\cite{Hawking:1982dh,Witten:1998zw} between the thermal
$AdS$ background and the $AdS$ black hole happens at $\beta_c =\pi
R/2$ $(r_0=R/2)$ above the temperature bound $\beta_c<\pi R/\sqrt{3}$
as seen in Eq.~\eqref{ENT41}. In the field theory side, the Hawking-Page
transition can be interpreted as the confinement/deconfinement
transition, since the free energy is the order parameter of the phase transition, which changes from being zero for the thermal $AdS$ in the planar limit
to being of order $N^{3/2}$ for the $AdS$ black hole. Note that
since the critical temperature does not depend on the orbifold but
depends on $R$, the critical temperature is not changed between the
mother theory and the daughter theory of the orbifold where we have
the same $AdS$ radius. In other words, the $AdS$ part is not affected, 
and this means that the Hawking-Page transition indeed occurs at
the same temperature. This result shows that in the field theory
side dual to the Eschenburg space, the critical point of the
confinement/deconfinement transition does not change under the
orbifold action, since symmetry of the orbifold is not broken in the
deconfinement phase. For the confinement phase corresponding to the
thermal $AdS$ case, on the other hand, we cannot discuss the
equivalence, since the free energy vanishes in the planar limit.

\section{EXPLANATION BY USING MIRROR SYMMETRY}

Mirror symmetry of the $U(N)_k\times U(N)_{-k}$ ABJM theory is
considered in Refs.~\cite{Jensen:2009xh,Hanada:2011zx} including a step of the mass deformation for fundamentals in the type-IIB string theory.
In this section, we use mirror symmetry in the type-IIB elliptic
D3 brane configuration to explain the $\mathbb{Z}_k$ orbifold
equivalence of the $\Nthree$ Chern-Simons matter theories
with flavor.
Mirror symmetry takes a theory with coupling $g_{YM}^2= O(1/N)$ to
a theory with coupling $g_{YM}^{\prime 2}\sim 1/g_{YM}^2=O(N)$.
We start with the original theory and after taking a mirror dual,
we analyze the orbifold of the mirror theory in the 't Hooft limit.
Remember that when $k$ is small, we should see the IR fixed points
of the original theory of the energy $E/\lambda_{3d}\sim
O(1/N^2)$.\footnote{When $k\sim N$, there are many massless states
in the mirror theory.} The IR fixed point actually describes the region where Yang-Mills terms decouple in the presence of the adjoint mass.

\begin{figure}[htbp]
   \begin{center}
     \includegraphics[height=4cm]{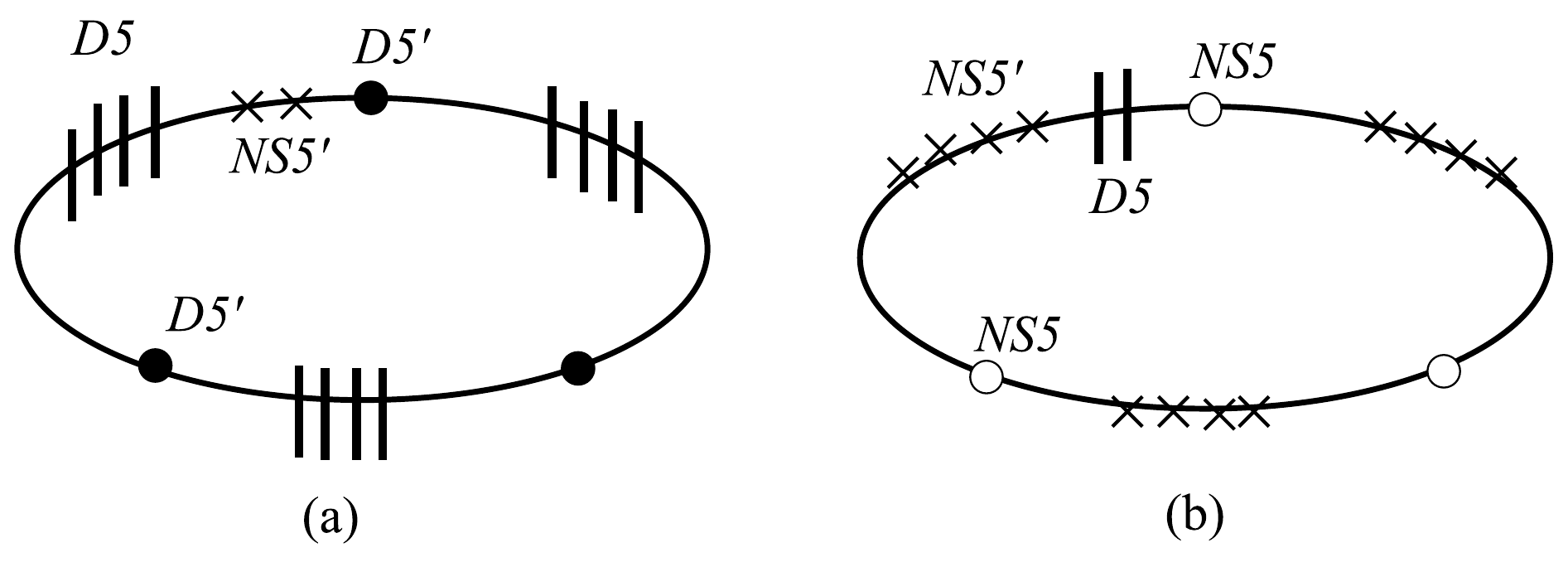}
   \end{center}
   \caption[mirror]{(a) The $\Ntwo$ type-IIB brane configuration of the original theory for $k=3$ and $N_F=12$. (b) The mirror dual of (a). }
\label{fig:mirror}
\end{figure}

To obtain the mirror to the $\Nthree$ Chern-Simons matter
theories with flavor, we start with the $\Ntwo$ type-IIB
brane configuration of the original theory  given by $N$ D3 branes,
$k$ D5$'$ branes, 2 NS5$'$, and $N_F$ D5 branes with different
orientations where we consider $N_F=kN_F'$ numbers of D5 for
convenience.
 They are given in Table I.
 \begin{equation}\label{D3a}
 \text{Table I: Type-IIB brane configuration of the original theory.}  \nonumber
 \end{equation}
\begin{equation}
\begin{array}{r|cccccccccccl}
\,\,  & x_0     & x_1   & x_2 & x_3  &
x_4 & x_5   & x_6   & x_7   & x_8   & x_9 & \, \nonumber\\
\hline N
\,\, \mbox{D3:}\,\,\,& \x &   \x  &  \x &   &  &   & \x &  &
&
&\,\nonumber\\
k\,\,  \mbox{D5$'$:}\,\,\,& \x &   \x  &  \x & \x  & \x  &   &  &  &
&\x &\,\nonumber\\ 2 \,\, \mbox{ NS5$'$:}\,\,\, & \x &  \x & \x  & \x  & \x
& \x  &   & & & &\, \nonumber \\
N_F\,\, \mbox{ D5:}\,\,\, & \x &  \x & \x  &   & 
&   &   &\x & \x&\x &\,
\end{array}
\end{equation}
Note that since there are two transverse directions for both D5 branes
and three transverse directions for both D5$'$ and NS5$'$, 
we can exchange them on the D3 branes without hitting each
other{~\cite{Hanany:1996ie}}.
5-branes are placed on D3 branes and aligned along $x_6$ in the
order of 2 NS5$'$, a D5$'$, $N_F'$ D5, a D5$'$, $N_F'$ D5$\dots$,
and $N_F'$ D5, where D5 and D5$'$ are placed to be symmetric in terms
of $\mathbb{Z}_k$ in the absence of 2 NS5$'$. See Fig.~\ref{fig:mirror}(a).

We consider the mirror duality by exchanging D5 branes with NS5 branes
and make D3 branes invariant.
The $\Ntwo$ type-IIB brane configuration of the mirror
theory is given by $N$ D3 branes, $k$ NS5, 2 D5 branes, and $N_F$
NS5$'$ with different orientations. They are given in Table II.
\begin{equation}
\text{Table II: Type-IIB brane configuration of the mirror theory. } \nonumber\end{equation}
\begin{equation}
\begin{array}{r|cccccccccccl}
\,\,  & x_0     & x_1   & x_2 & x_3  &
x_4 & x_5   & x_6   & x_7   & x_8   & x_9 & \, \nonumber\\
\hline N
\,\, \mbox{D3:}\,\,\,& \x &   \x  &  \x &   &  &   & \x &  &
&
&\,\nonumber\\
k\,\,  \mbox{NS5:}\,\,\,& \x &   \x  &  \x &   &   & \x   &  &\x  &\x
& &\,\nonumber\\ 2 \,\, \mbox{ D5:}\,\,\, & \x &  \x & \x  &   & 
&   &   &\x &\x &\x &\, \nonumber \\
N_F\,\, \mbox{ NS5$'$:}\,\,\, & \x &  \x & \x  & \x  & \x 
&  \x &   & & & &\,
\end{array}
\end{equation}
5 branes are placed on D3 branes and aligned along $x_6$ in the
order of 2 D5, a NS5, $N_F'$ NS5$'$, a NS5, $N_F'$ NS5$'$$\dots$,
and $N_F'$ NS5$'$, where NS5 and NS5$'$ are placed to be rotationally
$\mathbb{Z}_k$ symmetric  in the absence of 2 D5. See Fig.~\ref{fig:mirror}(b).

The gauge group of the above configuration becomes $[U(N)^2\times
U(N)^{N_F'-1}]^k$. In the aligned NS5 and NS5$'$ configurations,
we have the adjoint matter between two NS5$'$ branes with the same
direction, and we do not have the adjoint matter between the NS5 and
the NS5$'$~\cite{Elitzur:1997fh,Brunner:1998jr,Bergman:1999na},
because we have only $\Ntwo$ supersymmetry.
We consider the cross configuration where each D5 brane is on top
of a single NS5 brane. This configuration preserves $d = 3$, $\Ntwo$ supersymmetry and for a single NS5 brane, there are four copies of fundamental chiral
multiplets for the two gauge groups via the flavor doubling. The
global symmetry associated with these chiral multiplets is enhanced
from $U(1)^2$ to $U(1)^4$.

On the other hand, we can take the $\mathbb{Z}_k$ orbifold for the
mirror theory with $kN$ D3 branes, two D5, one NS5, and $N_F'$
NS5$'$. We consider the cross configuration where each D5 brane is
on top of a NS5. The gauge group of the mother theory is then
$\prod_{i=1}^{1+N_F'}U(kN)_i$.
The matter content consists of $N_F'-1$ chiral multiplets in the
adjoint $Y_i$ $(i=3,\dots,N_F'+1)$, $1+N_F'$ hypermultiplets
transforming in the bifundamental representation of $(i,i+1)$ gauge
groups $(A_{i,i+1},B_{i+1,i})$, four chiral multiplets transforming
in the fundamental representation under the first- and second-gauge
group $L_{(a)1},R_{(a)2}$, and four chiral multiplets transforming
in the antifundamental $\tilde{L}_{(a)1}, \tilde{R}_{(a)2}$ where
$a = 1, 2$. The $\Ntwo$ superpotential becomes
\begin{align}
S=\sum_{a=1}^2[\tilde{L}_{(a)1}A_{1,2}R_{(a)2}-\tilde{R}_{(a)2}B_{2,1}L_{(a)1}].
\end{align}

The $\mathbb{Z}_k$ orbifold projection is obtained from the element
of each gauge group $\prod_{i=1}^{1+N_F'}U(kN)_i$ and spans a
$\mathbb{Z}_k$ subgroup~\cite{Park:1999eb} as
\begin{align}
\gamma = \text{diag}(1_{\mathbf{N}},\ \omega 1_{\mathbf{N}},\ \omega^2 1_{\mathbf{N}},\ \dots\ ,\omega^{k-1} 1_{\mathbf{N}}),
\end{align}
where $1_{\mathbf{N}}$ is the $N\times N$ identity matrix and we
have defined the phase $\omega =e^{2\pi i/k}$. $k$ should be
relatively prime to $N_F'+1$, since otherwise the quiver diagram
is separated into many parts, as also observed in the case of orbifolds
of the ABJM theory~\cite{Terashima:2008ba}.

The quiver gauge theory is obtained from the $\prod_{i=1}^{1+N_F'}U(kN)_i$
theory by keeping the components that are invariant under the
orbifold projection as
\begin{align}
&V_i\to \gamma V_i\gamma^{-1},\quad Y_i\to \gamma Y_i\gamma^{-1}, \\
&A_{i,i+1}\to \omega \gamma A_{i,i+1}\gamma^{-1},\quad B_{i+1,i}\to \omega^{-1}\gamma B_{i+1,i}\gamma^{-1}, \\
&\tilde{L}_{(a)1}\to \tilde{L}_{(a)1}\gamma^{-1},\quad R_{(a)2}\to \omega^{-1}\gamma R_{(a)2}, \quad (a=1,2), \\
& L_{(a)1}\to \gamma L_{(a)1},\quad \tilde{R}_{(a)2}\to \omega \tilde{R}_{(a)2}\gamma^{-1} (a=1,2).
\end{align}

After the $\mathbb{Z}_k$ orbifold, the gauge group becomes
$[U(N)^2\times U(N)^{N_F'-1}]^k$ and realizes the mirror brane
configuration. The flavor fields couple to two nodes separated by
a NS5 brane. In addition, the $\mathbb{Z}_k$ symmetry of the daughter
theory is now seen as the symmetry rotating $1+N_F'$ units of the
nodes along the quiver diagram in the absence of the flavor. The
presence of the flavor breaks $\mathbb{Z}_k$ symmetry in the UV
of the daughter theory. However, according to Ref.~\cite{Jensen:2009xh},
mirror symmetry implies that in the deep IR of this gauge theory,
all 5-branes are gathered in the same position on D3 branes, and the
global symmetry $U(1)^4$ is enhanced to $U(k)\times U(k)\times
U(2)\times U(2)\times U(N_F)$ for both the original theory and the
mirror theory. Thus, we seem to recover $\mathbb{Z}_k$ symmetry in
the deep IR.

The flow to the IR fixed point described by the $\Nthree$
Chern-Simons matter theory is not directly given by the above brane
configurations. Though the field content is the same, we should
give a mass deformation for the fields coming from the D5 branes
and D5$'$ branes in the original $U(kN)^2$ theory. In the mirror
theory side $\prod_{i=1}^{1+N_F'}U(kN)_i$, the mass deformation
maps to some nonlocal deformation such as the monopole
operator~\cite{Hanada:2011zx}. It implies that the Lagrangian description of the mirror dual does not exist. However, it can be shown that large-$N$ orbifold equivalence can be proven by not using the Lagrangian description but instead using the brane configuration nonperturbatively.

\section{DISCUSSION}

In this paper, we showed two large-$N$ orbifold equivalences between $d=3$,
 $\Nthree$ and $4$  Chern-Simons matter theories. Here,
$\Nthree$ Chern-Simons matter theories include flavor.
We first analyzed the $\mathbb{Z}_p$ orbifold equivalence for the
orbifold changing the nodes of the gauge groups.
For the $\Nfour$ case, we found evidence that the $\mathbb{Z}_p$ orbifold
equivalence holds from the M theory limit to the weak coupling limit
by analyzing the gravity dual and the free theory on $S^1\times
S^2$. For the analysis in the free theory, we showed that the
free energy, the Polyakov loop VEV, and the critical temperature
of the phase transition agree with the relation expected in the
orbifold equivalence~\cite{Larsen:2007bm,Unsal:2007fb}.

{For the $\Nthree$ case with flavor, we can believe the equivalence
when flavor is aligned to reflect the $\mathbb{Z}_p$ symmetry in
the daughter theory. We showed that the $\mathbb{Z}_p$ equivalence holds in the
M-theory region using the gravity dual. When $N_F\ll N$, large-$N$ orbifold equivalence using the type-IIA string theory could be analyzed by introducing probe D6 branes corresponding to flavor without backreactions of them. Here, the $\mathbb{Z}_{2p}$ holonomy on the probe brane was used to specify the node coupling to fundamentals. That is, it implies that using the gravity dual, the Douglas-Moore orbifold projection is appropriate when we add small flavor degrees of freedom. Actually, the superconformal indices for the Chern-Simons matter theories with flavor 
 dual to an Eschenburg space were consistent with the gravity dual including the contributions from D6 branes wrapping $\mathbb{RP}^3$~\cite{Cheon:2011th}.
 
 For the case of backreacted flavor, the dilation or the coefficient of the M circle depends on the internal coordinates~\cite{Gaiotto:2009tk} under the dimensional reduction to the type-IIA superstring theory. It also means that when $k\neq 1$, there are fixed points at the $\mathbb{Z}_k$ singularity. 
 We leave large-$N$ equivalence in this type-IIA string theory for future work. It will be interesting to analyze the orbifold equivalence of the
Chern-Simons matter theory with flavor in the weak coupling limit
where we should include finite $\lambda$ corrections to describe
the first-order phase transition instead of the third-order phase
transition  for zero 't Hooft coupling.}
 
Secondly, we analyzed the $\mathbb{Z}_{k}$ orbifold equivalence by
changing Chern-Simons levels in the M-theory region. {We confirmed large-$N$ 
equivalence by computing the BPS monopole operators sensitive to the $\mathbb{Z}_{k}$ projection (see also Ref.~\cite{Hanada:2011yz}). It will be interesting to apply the equivalence between the BPS monopole operators for other Chern-Simons matter theories. We also showed that the critical temperature of Hawking-Page transition does not change, since the $AdS$ part is not affected by the orbifold and the Bekenstein-Hawking entropy behaves as expected in large-$N$ orbifold equivalence. In the M-theory limit, the entropy Eq.~\eqref{ENT417}  was also consistent with the planar dominance outside the planar limit~\cite{Azeyanagi:2012xj}. It implies that large-$N$ planar equivalence holds even outside the 't Hooft limit when there exists a classical gravity dual. It is known, however, that the large-$N$ equivalence is broken when $1/N$ corrections coming from the nonplanar diagram are included.}
In the $\mathbb{Z}_k$ equivalence, the change of Chern-Simons levels
can be interpreted as a change of the number of D5 branes in the
mirror theory side of the type-IIB elliptic D3 brane configuration.

In Ref.~\cite{Hanada:2011yz}, the $\mathbb{Z}_k$ equivalence of
Chern-Simons matter theory was also confirmed from the free energy
computed by using the localization
method~\cite{Kapustin:2009kz,Drukker:2010nc,Fuji:2011km,
Marino:2011eh,Hanada:2012si}. 
{For unquenched flavor case, since the behavior of the free
energy~\cite{Herzog:2010hf,Santamaria:2010dm} was the same as that
derived from gravity dual $N^{3/2}/\sqrt{vol(\mathcal{M}_7})$, it was consistent with the $\mathbb{Z}_k$ orbifold equivalence
between the $U(kN)_1\times U(kN)_{-1}$ theory with $N_F$ flavor and the
$U(N)_k\times U(N)_{-k}$ theory with $kN_F$ flavor.} 

It will also be interesting to analyze large-$N$ equivalence between the Kaluza-Klein spectra on the Eschenburg space and that of its orbifold to understand the $\mathbb{Z}_k$ orbifold projection with the fixed point in the gravity dual. Actually,
for $(t_1,t_2,t_3)=(1,1,1)$, $S_7^{(1,1)}(1,1,1)=N(1,1)$ and its
Kaluza-Klein spectra are known in Refs.~\cite{Termonia:1999cs,Fre':1999xp,Billo:2000zr}. In the recent work Ref.~\cite{Cheon:2011th}, moreover, the superconformal indices for the $\mathcal{N}=3$ Chern-Simons matter theories with flavor dual to $N^{010}/\mathbb{Z}_k$ are realized by counting the Kaluza-Klein spectrum of the dual supergravity.

\paragraph*{Acknowledgements} 
{We would like to thank A. Armoni, S.~Cremonesi, S.~Janiszewski, S.~Kachru, A.~Karch, E.~Silverstein, S.~Sugimoto, T.~Takayanagi, and T.~Watari for helpful discussions and comments. We would like to thank C. Hoyos and L. Yaffe for collaboration in the initial stage of this project and for helpful and valuable comments. 
M. F. is in part supported by JSPS Postdoctoral Fellowship and partly by JSPS Grant-in-Aid for JSPS Fellows No. 25-4348. This work was supported by the World Premier International Research Center Initiative (WPI Initiative), MEXT, Japan.}

\end{document}